\documentstyle[11pt,rotating,epsfig]{article}

\newcommand{\unit}[1]{\mbox{\ #1}}

\newcommand{\comment}[1]{}
\newcommand{\str}{\rule{0ex}{2.7ex}}

\newcommand{\mcal}{{\cal M}}
\newcommand{\be}{\begin{equation}}
\newcommand{\ee}{\end{equation}}

\renewcommand{\Re}{\mbox{Re}}

\newcommand{\bildchen}[3]{%
\begin{center}
\begin{sideways}
\makebox{\hspace*{5ex} \Large $\displaystyle #2$}
\end{sideways}
\mbox{{\epsfig{figure=#1,width=13.cm,%
bbllx=1.8cm,bblly=9.2cm,bburx=20.cm,bbury=19.cm}}}
\end{center}
\begin{flushright}
   {\Large $\displaystyle #3$ \hspace*{5ex}}
\end{flushright}}

\begin{document}
\everymath={\displaystyle}
\thispagestyle{empty}
\vspace*{-2mm}
\thispagestyle{empty}
\noindent
\hfill LNF preprint\\
\mbox{}
\hfill hep-ph/9412267\\
\mbox{}
\hfill November 1994 \\
\vspace{0.5cm}
\begin{center}
  \begin{Large}
  \begin{bf}
RADIATIVE CORRECTIONS\\
TO $K_{l2}$ DECAYS\\
\mbox{\ }   \\
  \end{bf}
  \end{Large}
  \vspace{0.8cm}
  \begin{large}
    Markus Finkemeier\\[5mm]
    INFN - Laboratori Nazionali di Frascati\\
	P.O.~Box 13\\
	00044 Frascati (Roma)\\
	Italy\\
  \end{large}
  \vspace{4.5cm}
  {\bf Abstract}
\end{center}
\begin{quotation}
\noindent
We consider $K_{l2}$ decays to the order $O(\alpha)$. We perform a
matching calculation, using a specific model with vector meson dominance
in the long distance part and the parton model in the short distance
part. By considering the dependence on the matching scale and on the
hadronic parameters, and by comparing with the leading model independent
predictions, we scrutinize the model dependence of the results.
\end{quotation}
\vspace*{2cm}
\begin{center}
Contribution to the Second Edition of the Daphne Physics Handbook,\\
edited by L. Maiani, G. Pancheri and N. Paver\\[2cm]
\end{center}
\newpage
%
\section{Lowest Order Prediction}
{}From the theoretical point of view, $K_{l2}$ decays
\begin{eqnarray}
   K^+ & \to & \mu^+ \nu_\mu
\nonumber \\
   K^+ & \to & e^+ \nu_e
\end{eqnarray}
constitute the simplest semileptonic decay modes of the kaon.
Their decay rates depend on the matrix element
of the axial weak hadronic current
\be
   < 0 | A^\mu(0) | K^+(p)> = i \sqrt{2} f_K p^\mu
\ee
and, neglecting electromagnetic corrections, the decay rates are given by
\be
   \Gamma(K^+ \to l^+ \nu_l) =
   \frac{G_F^2 V_{us}^2}{4 \pi} f_K^2 m_K m_l^2
   \left(1 - \frac{m_l^2}{m_K^2} \right)^2
\ee
Using \cite{RPP94}
\begin{eqnarray}
   \Gamma(K \to\mu\nu_\mu(\gamma)) & = & (3.38 \pm 0.01)\cdot 10^{-14}\,
  \mbox{MeV}
\nonumber \\
	|V_{us}| & = & 0.2205 \pm 0.0018
\end{eqnarray}
we extract the kaon decay constant to the order $O(\alpha^0)$
\be
   f_K^{(0)} = (113.4 \pm 0.9) \, \mbox{MeV}
\ee
where the error is due to the uncertainty in $V_{us}$.

Of course, $f_K$ and  $f_\pi$, which is extracted
from $\Gamma(\pi\to\mu\nu_\mu)$ in an analogous way, are important input
parameters for chiral perturbation theory \cite{Gas84,kapintro}.
It is therefore important to understand the effect
of $O(\alpha)$ radiative
corrections on their extraction \cite{Hol90,Mar91,Mar93}.

Whereas $f_\pi$ and $f_K$ can not be predicted without being able to
solve the non-perturbative sector of QCD, we can predict the ratio $R_K$
of the electronic and muonic decay modes of the kaon.
At the order $O(\alpha^0)$, we obtain
\be
   R_K^{(0)} := \frac{\Gamma(K\to e \nu_e )}
               {\Gamma(K\to \mu \nu_\mu )}
   = \frac{m_e^2}{m_\mu^2} \left( \frac{m_K^2 - m_e^2}
     {m_K^2-m_\mu^2} \right)^2
 = 2.60 \cdot 10^{-5}
\ee
There are two points which should be emphasized.
First, note that to the $O(\alpha^0)$, strong interaction effects, which
we have
parameterized by $f_K$, cancel completely in this ratio, and it is
predicted in terms of the precisely measured particle masses only.
As we will show, this cancelation of strong interaction effects occurs
also to a large extent for the $O(\alpha)$ radiative corrections,
allowing for a very precise prediction of $R_K$.
Secondly, we observe the famous dynamical helicity suppression of the
electronic mode, which is of course due to the fact that the weak
interaction is mediated by a spin-1 particle. (Note that the helicity
suppression does not only occur for a $V-A$ interaction, but for any
$(V,A)$ structure.)
Therefore the ratio $R_K$ allows for an important precision test of the
standard model. Although at least at present \cite{RPP94}
the experimental precision in pion decays is better than in kaon decays,
and so the measurement of the corresponding ratio $R_\pi$ is more
precise, the study of $K_{l2}$ is of equal importance, because due to
the larger kaon mass, the effects of non-standard physics might be
enhanced in $R_K$ by a factor of $m_K/m_\pi$.
%
\section{Structure of the Radiative Corrections}
%
At the order $O(\alpha)$ one has to consider inclusive decay rates into
final states with or without an additional photon,
\be
  \Gamma(K \to l \nu_l) + \Gamma(K \to l \nu_l \gamma)
\ee
because of the infra-red divergences in the virtual and soft photonic
corrections, which cancel in the inclusive sum only.
At this point, one can either to integrate over all photons, soft and
hard ones, or to include only soft photons, using some type of an
upper limit on the photon
energy.

So we have to consider the Born amplitude $\mcal_0$,
the amplitude of 1-loop virtual photonic corrections
$\delta \mcal_v$, and the amplitude $\delta \mcal_r$ of the
radiative decay:
\begin{eqnarray}
   \mcal_0(K\to l \nu_l)
\nonumber \\
   \delta \mcal_v(K\to l \nu_l)
\nonumber \\
  \delta \mcal_r(K\to l \nu_l \gamma) & = &\mcal_{IB} + \mcal_{SD}
\end{eqnarray}
The radiative decay $K\to l\nu_l \gamma$ is discussed in detail in this
handbook in the section on semileptonic kaon decays \cite{kap2}.
Note
that $\mcal_{IB}$ is of the $O(P^2)$, whereas $\mcal_{SD}$ starts at the
order $O(P^4)$. The internal bremsstrahlung amplitude $\mcal_{IB}$ is
identical to what is obtained assuming an effective point like kaon, and so
$\mcal_{SD}$ includes all the hadronic structure dependent (``SD'')
effects.
These amplitudes give rise to the Born level decay rate $\Gamma_0$, to
the virtual correction $\delta \Gamma_v$ of the decay rate and to the
contributions $\Gamma_{IB}$, $\Gamma_{INT}$ and $\Gamma_{SD}$ to the
radiative decay:
\begin{eqnarray}
   \Gamma_0 & \propto & | \mcal_0| ^2
\nonumber \\
   \delta \Gamma_v & \propto & 2 \Re(\mcal_0 \delta \mcal_v^\star)
\nonumber \\
   \Gamma_{IB} & \propto & | \mcal_{IB}|^2
\nonumber \\
   \Gamma_{INT} & \propto & 2 \Re(\mcal_{IB} \mcal_{SD}^\star )
\nonumber \\
   \Gamma_{SD} & \propto & | \mcal_{SD}|^2
\end{eqnarray}
$\Gamma_{IB}$ is proportional to $\Gamma_0$ and therefore vanishes for
$m_l \to 0$. $\Gamma_{SD}$, on the other hand, is not helicity
suppressed and becomes phase space enhanced for $m_l\to 0$.
So in matter of fact, for the electronic decay mode of the kaon, where
$m_e \ll m_K$, the amplitude for structure dependent radiation is of the
same order of magnitude as the Born decay rate (see also \cite{kap2}):
\be
   \Gamma_{SD}(K\to e \nu_e \gamma) \approx \Gamma_0(K\to e \nu_e)
\ee
Therefore we adopt the following convention, concerning the question of
hard photons, which is more or less identical to the one used
in the extraction of the experimental data.
We include the full $\Gamma_{IB}$ contribution, which is
strongly dominated by very soft photons, but we exclude completely the
structure dependent $\Gamma_{SD}+ \Gamma_{INT}$ contribution,
which is strongly dominated by hard photons.

Thus we calculate
\be
   \Gamma(K\to l\nu_l (\gamma)) = \Gamma_0 + \delta \Gamma
\ee
where the radiative correction $\delta \Gamma$ is defined by
\be
   \delta \Gamma := \delta \Gamma_v + \Gamma_{IB}
\ee
Note that while for the electronic mode the excluded structure dependent
contribution is extremely large, it is completely negligible
for $K\to\mu\nu_\mu\gamma$.

Of course it is not possible to tell whether a radiated photon
is due to internal bremsstrahlung or to structure dependent radiation.
However, if some small upper limit on the photon energy is used,
the measured rate of $K \to
e \nu_e (\gamma)$ will include only a very small SD + INT background,
and only very little of the IB part will have been discarded. Using
the predicted differential distributions, which are given in detail in
\cite{kap2}, the SD +INT
background can be subtracted and the missing IB part added. Because of
the smallness of this correction, it does not give rise to
any important uncertainties.

The particle data book \cite{RPP94} quotes numbers for $K \to e \nu_e$,
without stating clearly how much $K \to e \nu \gamma$ is
included in these numbers. However,
from reading the original papers such as \cite{Hea75}, we believe that
our convention comes close to the procedure applied
in the extraction of the experimental data.
With increasing experimental
precision, it will become important that experimentalists state very
clearly which corrections with respect to the radiative decay $K \to e
\nu_e \gamma$ have been applied, and we suggest to use the approach
described above.

Let us now discuss the virtual corrections is some more detail.
Notice that we are discussing processes
at a scale much smaller than $m_W^2$. Therefore we can use
an effective local interaction, cut off at $m_W^2$ by
the $W$ propagator. In fact, it has been shown by Sirlin \cite{Sir78}, that
to the order $\alpha G_F$, the radiative corrections within the full
standard model are identical to the photonic corrections calculated
with a local $V-A$ interaction and an ultra-violet cut-off equal to
$m_Z$. In the proof, terms of the order $\alpha_s \alpha G_F$ and $G_F
\alpha m^2/m_W^2$ are neglected, where $m$ is the mass of a quark or of
an external particle.

In order to be able to calculate the virtual corrections, one separates
the loop integration $k_E^2  = 0 \cdots m_Z^2$ into two
parts, using a matching scale $\mu_{cut}$ of the order of $1\,
\mbox{GeV}$ \cite{Sir72,Hol90,Dec94}.
In the long distance part,
$k_E^2 = 0 \cdots \mu_{cut}^2$, mesons are the relevant degrees of
freedom.
A reasonable first approximation for the long distance part is in fact
a model using an effective point like kaon. This has then to be modified to
account for hadronic structure effects.
In the short distance part,
$k_E^2 =  \mu_{cut}^2 \cdots m_Z^2$, on the other hand, quarks are the
relevant degrees of freedom, and one has to
consider 1-loop photonic corrections to the effective four-point
operator ${\cal A}_0 = [\bar{u}_\nu \gamma^\mu \gamma_- u_l]
[\bar{u}_s \gamma_\mu \gamma_- u_u]$.

Thus after adding the rate for internal bremsstrahlung,  the
total radiative correction can be written as the sum of three terms,
\be
    \frac{\delta \Gamma}{\Gamma_0} (K\to l \nu_l (\gamma))
   = \left(\frac{\delta \Gamma}{\Gamma_0}\right)_{PM}
    + \left(\frac{\delta \Gamma}{\Gamma_0}\right)_{HSD}
    + \left(\frac{\delta \Gamma}{\Gamma_0}\right)_{s.d.}
\ee
the correction obtained with an effective point meson (PM), the hadronic
structure dependent correction (HSD), and the short distance (s.d.)
correction.

The point meson correction has been calculated long ago by Kinoshita
\cite{Kin59},
\be
  \left(\frac{\delta \Gamma}{\Gamma_0}\right)_{PM} =
  \frac{\alpha}{2 \pi} \left\{ \frac{3}{2} \ln \frac{m_K^2}{\mu_{cut}^2}
  + 6 \ln \frac{m_l}{m_K} + \frac{11}{4} - \frac{2}{3} \pi^2 +
   f\left( \frac{m_l}{m_K} \right) \right\}
\ee
where we have defined $f(r_l)$ in such a way that $f(0)=0$:
\begin{eqnarray}
   f(r_l) & = & 4 \left(\frac{1+r_l^2}{1-r_l^2} \ln r_l
   - 1 \right) \ln(1-r_l^2)
   - \frac{r_l^2 (8 - 5r_l^2)}{(1-r_l^2)^2}\ln r_l
\nonumber \\ \nonumber \\
   & & + 4 \frac{1 + r_l^2}{1-r_l^2} \mbox{Li}_2(r_l^2)
   - \frac{r_l^2}{1-r_l^2}
   \left( \frac{3}{2} + \frac{4}{3} \pi^2 \right)
\end{eqnarray}
$\mbox{Li}_2$ denotes the dilogarithmic  function
\be
   \mbox{Li}_2(x) = - \int_0^x dt \, \frac{\ln(1-t)}{t}
\ee

The leading contribution to the short distance correction in the limit
$m_Z^2\to\infty$ of a large $Z$ boson mass has been given by Sirlin
\cite{Sir82}
\be
  \left(\frac{\delta \Gamma}{\Gamma_0}\right)_{s.d} =
  \frac{2 \alpha}{\pi} \ln \frac{m_Z}{\mu_{cut}} + \cdots
\ee

Finally, following \cite{Mar93}, we classify the hadronic structure
dependent corrections in terms of the dependence on the lepton mass,
\be \label{eqn18}
    \left(\frac{\delta \Gamma}{\Gamma_0}\right)_{HSD}
  = - \frac{\alpha}{\pi} \left\{ C_1 + C_2 \frac{m_l^2}{m_\rho^2}
   \ln \frac{m_\rho^2}{m_l^2} + C_{lms} \frac{m_K^2}{m_\rho^2}
   \ln \frac{m_\rho^2}{m_l^2} + C_3 \frac{m_l^2}{m_\rho^2}
   + \cdots \right\}
\ee
where the coefficients $C_i$ are independent on the lepton mass.
The dots $\cdots$ indicate more suppressed lepton mass dependent terms.
Note
that in \cite{Mar93} the term proportional to $C_{lms}$ is not
considered. It diverges in the limit $m_l\to 0$ and therefore gives rise
to a lepton mass singularity in the radiative correction. Note that such
a mass singularity is allowed in spite of the Kinoshita-Lee-Nauenburg
\cite{Kin59,Lee64} theorem, because the Born amplitude is proportional to
$m_l^2$.
According to a theorem by Marciano and Sirlin \cite{Sir76},
however, the hadronic structure
dependent effects on the coefficient of the lepton mass singularity
cancel between the virtual $\delta \Gamma_v$
and the real photonic corrections $\delta \Gamma_r$,
leaving the point meson result unchanged, if we integrate the real
photons over the full phase space.
However, this cancelation only
works if we include the full $\Gamma_{SD}+\Gamma_{INT}$ contribution,
which we do not do.

There is no way to calculate $C_1$ and $C_3$ in a model independent way.
The coefficients $C_2$ and $C_{lms}$, on the other hand, are model
independent and have been determined in \cite{Ter73}

It should be emphasized at this point that the extraction of the
pseudoscalar decay constants $f_M$ beyond the order
$O(\alpha^0)$ is not unambiguous. One could by definition include part
of the radiative correction into the $f_M$, e.g.\ the short distance
part or the model dependent, lepton mass independent contribution from
$C_1$. However, we choose to factor out all $O(\alpha)$ effects from the
decay constants. This definition does not agree with the one
used by Holstein \cite{Hol90}, who absorbs process dependent
terms proportional to $\ln
(m_\pi/m_\mu)$ into $f_\pi$. See also the discussion in
\cite{Mar91,Mar93} about this point.
These ambiguities are of course due to the fact that $f_\pi$ and $f_K$
are no observables. The ambiguities cancel, however,
in the ratios of the decay rates
\be
   R_M := \frac{\Gamma(M\to e \nu_e (\gamma ))}
               {\Gamma(M\to \mu \nu_\mu (\gamma))}
   = \frac{m_e^2}{m_\mu^2} \left( \frac{m_M^2 - m_e^2}
     {m_M^2-m_\mu^2} \right)^2 \Big( 1 + \delta R_M \Big)
\ee
and the radiative correction $\delta R_K$ is defined unambiguously and
can in fact be predicted very precisely.

In the extraction of the decay constants $f_\pi$, $f_K$ and in the
calculation of the ratio $R_K$ one has to decide which matching scale to
use and what to do about the model dependent corrections.

In \cite{Mar93}, the authors consider $\pi_{l2}$ decays. Their approach
is the following. They include only the leading model
independent terms in
the long distance part and use a rather low matching scale of
$\mu_{cut} = m_\rho$. To extract $f_\pi$ from $\Gamma
(\pi\to\mu\nu_\mu(\gamma))$, they estimate $C_1$ very roughly by
equating it with the effect of varying the cut-off by a factor of two.
In the short distance correction, they include the also leading
$O(\alpha^n)$ corrections by using the renormalization group and the
leading logarithm of the order $O(\alpha \alpha_s)$.

In the ratio $R_\pi$, $C_1$ and all the short distance corrections,
which the authors have considered, cancel because they are independent
on the lepton mass. For the unknown coefficient $C_3$, which then
is the main source of uncertainty,
they consider a conservative range of $C_3 = 0\pm 10$.

We will use a somewhat different approach. For the long distance part we
use a phenomenological model, which includes vector ($\rho$,
$\omega$, $\Phi$, $K^\star$, \dots) and axial vector ($K_1$) resonances
as explicit degrees of freedom. This allows us to push the matching
scale up to $\mu_{cut} = 1 \dots 2 \, \mbox{GeV}$, rendering the
calculation of the short distance part more reliable. In the short
distance correction, which we calculate using the parton model,
we include small lepton mass dependent corrections which do not cancel
in $R_K$.
Within our model, we obtain an error estimate based on the dependence
on the matching
scale and on the hadronic parameters.
We then compare the predictions from our
model with the leading model independent corrections.
This allows us to obtain an
essentially model independent prediction by taking the
results from our model as central values and by taking the full
difference between our model and the model independent corrections as
uncertainty. We will be able to show in this way that the ratio $R_K$
can be predicted with an uncertainty of the order of $4 \cdot 10^{-4}$.
So indeed $K_{l2}$ decays offer the possibility of an important low
energy precision test of the standard model.
%
\section{Parametrization of the Amplitudes}
%
We will be brief here an refer to \cite{Dec94,Fin94} for details.

We separate the loop
integration into a long and a short distance part by splitting the
photon propagator
\begin{eqnarray}
   \frac{1}{k^2 - \lambda^2} & = &
   \underbrace{\frac{1}{k^2 - \lambda^2}\frac{\mu_{cut}^2}
   {\mu_{cut}^2 - k^2}}_{\mbox{``long distance''}}
   \quad
   + \underbrace{\frac{1}{k^2 - \mu_{cut}^2}}_{\mbox{``short distance''}}
\end{eqnarray}
using a matching scale $\mu_{cut} = (0.75 \dots 3) \, \unit{GeV}$.

The long distance part, involving a regulated photon propagator,
is calculated using a phenomenological model where mesons are the relevant
degrees of freedom. The short distance part, involving a massive
photon propagator
is calculated using the parton model.

To calculate the long distance corrections, we start from the amplitudes
obtained with an effective point like meson. These amplitudes are  good
approximations for very small momentum transfers only. Consider for
example the amplitude $V^\mu$ for the coupling of a photon to two pions.
In the point meson (P.M.) approximation it is given by
\be
   V^\mu(\pi^+(p) \pi^-(p') \to \gamma)^{(P.M.)} = i e (p-p')^\mu
\ee
However, this coupling {\em defines} the electromagnetic form factor
$F_\pi$ of the pion via
\be
   V^\mu(\pi^+(p) \pi^-(p') \to \gamma) =: i e F_\pi[(p+p')^2]
      (p-p')^\mu
\ee
Therefore we should modify the effective point
pion diagrams by multiplying
this coupling by $F_\pi$. This modification in turn determines by gauge
invariance the appropriate modification of the weak-electromagnetic
seagull coupling $\pi \gamma W$.

Analogously in the kaonic case, we modify the point kaon coupling by
multiplying it by $F_K$:
\be
   V^\mu(K^+(p) K^-(p') \to \gamma) \longrightarrow i e F_K[(p+p')^2]
      (p-p')^\mu
\ee
In \cite{Dec94} a parameterization of  $F_K$ with a simple $\rho$
dominance
\be
   F_K(t) = \mbox{BW}_\rho(t)
\ee
was used.
However, this assumes exact $SU(3)$ flavour symmetry,
$m_\rho = m_\omega = m_\Phi$. We will now drop this assumption. Thus we
have to consider the relative contributions of the $\rho$, the $\omega$
and the $\Phi$ to the form factor $F_K$. Assuming ideal mixing, $\Phi =
(s \bar{s})$, we obtain
\be
   F_K(t) = \frac{1}{2} \mbox{BW}_\rho(t)
   + \frac{1}{6} \mbox{BW}_\omega(t)
   + \frac{1}{3} \mbox{BW}_\Phi(t)
\ee
which we will use in the present paper.

In addition to the modified effective point meson diagrams, there are loop
diagrams which are obtained from the hadronic structure dependent
radiation (SD) by contracting the emitted photon with the lepton. If
$k^2$ is small, where $k$ is the momentum of the virtual photon, the
relevant form factors $H_V$ and $H_A$ in these hadronic structure
dependent loops will obviously be identical to $F_V$ and $F_A$
determining the radiative decay. However, they can additionally depend
on $k^2$.

For $k^2 = 0$ (on-shell photon), we use the following ansatz
\begin{eqnarray}
   F_V^{(K)}(t) =  F_V^{(K)}(0) \mbox{BW}_{K^\star}(t)
\nonumber \\
   F_A^{(K)}(t) = F_A^{(K)}(0) \mbox{BW}_{K_1}(t)
\end{eqnarray}
where
\begin{eqnarray}
   F_V^{(K)}(0) & = & 0.0955
\nonumber \\
   F_A^{(K)}(0) & = &  0.0525 \pm 0.010
\end{eqnarray}
Here $F_V^{(K)}(0)$ has been obtained from flavour symmetry and the
anomaly and $F_A^{(K)}(0)$ from this value for $F_V^{(K)}(0)$
and the measurement of the sum \cite{RPP94}.
This value is slightly higher than $F_A^{(K)}(0) = 0.0410$ which
corresponds to the $O(P^4)$ prediction of chiral perturbation theory
\cite{kap2}, which will, however, be subject to higher order flavour
symmetry breaking corrections.

$\mbox{BW}_X(t)$ denotes a normalized Breit-Wigner amplitude
\be
   \mbox{BW}_X(t) = \frac{m_X^2}{m_X^2 -t}
\ee
The virtual corrections are calculated in the Euclidean $k^2$ region,
and so we will not include an imaginary part of the mass.

For $k^2 \neq 0$, we adopt the following ansatz for the momentum
dependence, which is based on a double vector meson dominance:
\begin{eqnarray}
   H_V^{(K)}(k,p) = \mbox{BW}_V(k^2) F_V^{(K)}[(k-p)^2]
\nonumber \\
   H_A^{(K)}(k,p) = \mbox{BW}_V(k^2) F_A^{(K)}[(k-p)^2]
\end{eqnarray}
where $p$ is the momentum of the decaying kaon.
Note that here we assume $m_V = m_\rho = m_\omega = m_\Phi$.
We do not calculate the
$SU(3)$ flavour symmetry breaking effects. This will be justified
below by the observation that
the dependence of the result on $m_V$ is very small.

In order to obtain the short distance corrections,
we calculate the one-loop corrections $\delta {\cal A}$
to the operator
${\cal A}_0 = [\bar{u}_\nu \gamma^\mu \gamma_- u_l] \,
  [\bar{u}_s \gamma_\mu \gamma_- u_u]$.
Neglecting all masses except for $m_l$ and $\mu_{cut}$, we obtain
\be
   \left(\frac{\delta \Gamma}{\Gamma_0} \right)_{short\,\, dist.} \approx
   \frac{2 \alpha}{\pi} \frac{1}{m_l^2 - \mu_{cut}^2}
   \left(m_l^2 \ln \frac{m_Z}{m_l} - \mu_{cut}^2 \ln \frac{m_Z}
   {\mu_{cut}} \right)
\ee
which has to be compared to Sirlin's logarithm $2 \alpha/\pi \ln(m_Z /
\mu_{cut})$ \cite{Sir82}. Of course, if we do not neglect $m_\mu$, we should
also
take the meson mass $m_M$ into account. However, the contributions depending on
$m_M$
cancel in the ratios $R_M$, whereas in the radiative correction to the decay
rates themselves, $\mu_{cut}$ dominates anyway.

Note that for this leading logarithm, the correction to the quark level
short distance amplitude $\delta {\cal A}$ is proportional
to the Born amplitude ${\cal A}_0$. Thus the same logarithm is involved
in the corrections to the hadronic amplitude without any model
dependence resulting from hadronization.

While in the correction to the individual decay rates $M\to l \nu_l
(\gamma)$ this leading logarithm dominates the short distance
correction, it depends only very little on the lepton mass and thus
cancels almost completely in the ratios $\delta R_M$.
Therefore in the case of these ratios we go beyond the leading logarithm
and calculate the full one-loop short distance correction.

The complete result for $\delta {\cal A}$
is no longer proportional to the Born amplitude
${\cal A}_0$, and furthermore
it depends on the relative momentum of the two
quarks. Therefore we project onto the $J^P = 0^-$ component and
integrate over the relative momentum $u \times p$ of the quarks in the infinite
momentum frame ($u = -1 \dots +1$).
The result can
be written in the form
\be \label{eqn23}
   \Big( \delta R_K\Big)_{short\,\, dist.} =
   \frac{3}{2 f_K} \int_{-1}^{+1} du \, \Phi_K(u) r_K(u)
\ee
Here $\Phi_K(u)$ is an unknown parton distribution function
(kaon wave function), whereas $r_K(u)$ is calculated from the short
distance diagrams for arbitrary $u$. We find, however, that
$r_K(u)$ depends only very little on $u$, and we can approximate it
by its values at $u=0$, where the wave is presumably peaked:
\be \label{eqn24}
   \Big( \delta R_K \Big)_{short\,\, dist.} \approx
   r_K(u=0) \frac{3}{2 f_K} \int_{-1}^{+1} du \, \Phi(u) = r_K(0)
\ee
where the last equation follows from the Brand-Preparata
 sum rule \cite{Bra70}.
%
\section{Numerical Results}
%
Adding up long and short distance corrections, we obtain the full
radiative correction. This depends on the choice of the matching scale
$\mu_{cut}$ and on the hadronic parameters.

\begin{figure}
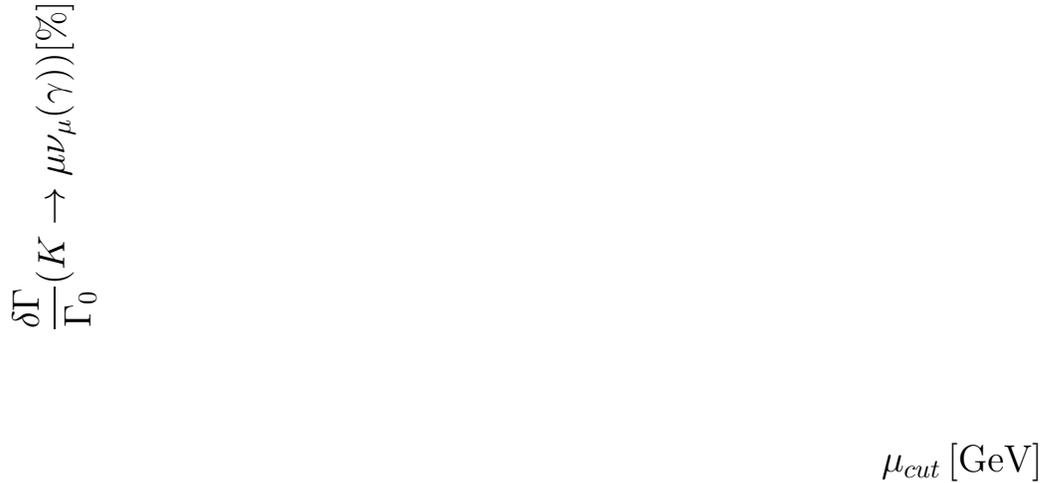

\caption{Radiative correction to $\Gamma(K\to\mu\nu_\mu)$, using
different choices for the hadronic parameters: Standard choice
(I, solid) and variations (II, dashed and III, dotted)}
\label{fig3}
\bildchen{fig2.ps}
{\frac{\delta \Gamma}{\Gamma_0}(K\to\mu\nu_\mu(\gamma)) [\%]}
{\mu_{cut}\,[\mbox{GeV}]}
\end{figure}
In Fig.~\ref{fig3} we display the correction to the decay rate
$\Gamma(K\to\mu\nu_\mu(\gamma))$
in variation with $\mu_{cut}$, using
three different choices for the hadronic parameters. The solid line (I)
corresponds to the central values given above. The dashed (II) and the
dotted (III) lines are obtained by varying the hadronic parameters,
viz.\ $F_A(0)$ and the relative contributions of higher radial excitations
in
$F_K$ and $H_V$.
Note that the matching scale dependence above $\mu_{cut} = 1 \,
\mbox{GeV}$ is very moderate,
indicating that our
phenomenological model for the long distance part is indeed rather
reasonable.

Choosing $\mu_{cut} = 1.5 \unit{GeV}$ as a
central value for
we find from Fig.~\ref{fig3} the following $O(\alpha)$ correction to
the
decay rate
\begin{eqnarray}
   \frac{\delta \Gamma}{\Gamma_0}(K\to\mu\nu_\mu(\gamma))
   & = &(1.23 \pm 0.13 \pm 0.02) \%
   + O(\alpha^2) + O(\alpha \alpha_s)
\nonumber\\
\nonumber\\
  & = & (1.23 \pm 0.13) \% + \cdots
\end{eqnarray}
The first error given ($0.13 \%$) is the matching uncertainty, obtained by
varying
$\mu_{cut}$ by a factor of two ($0.75 \dots 3 \, \mbox{GeV}$), the second
one ($0.02\%$) is the uncertainty from the hadronic parameters.

A few comments are in order:
\begin{enumerate}
\item As we have already states above,
this radiative correction is not a
physical
observable and therefore is  not defined unambiguously. The definition we
have adopted is to include {\em all} $O(\alpha)$
corrections in  the number $1.23\% $ given. Part of this might be
absorbed into $f_K$ by definition, but we choose not to do so.
\item
We employ $m_Z$ as an ultra-violet cut-off for the short
distance corrections, according to the general theorems of Sirlin
\cite{Sir78} on short-distance electroweak corrections to semileptonic
processes. But this implies that there is an arbitrariness of the order
of $\alpha/(2 \pi) \times O(1) \approx 0.1\% $ in the definition of the
radiative
correction, because a change of the cut-off scheme would induce a change
of the result of this order.
Note that the error of the $O(\alpha)$
correction which we have determined
is of the same order of magnitude as this inherent
ambiguity.
\item
The error $\pm 0.13 \%$ quoted above is the uncertainty of the
$O(\alpha)$ correction only. In \cite{Mar93} the authors have summed up
the leading $O(\alpha^n)$ corrections
for the dominant contribution $2 \alpha
/\pi\, ln (m_Z/\mu_{cut}$) in the short distance part, using the
renormalization group. This leads to an enhancement of the short distance
correction of $0.13\%$. Furthermore they considered the leading QCD short
distance correction, which decreases the short distance part by
$-0.03\%$. Similar $O(\alpha^n)$ effects should be considered in the long
distance part.
\end{enumerate}
Taking into account these higher order short distance corrections, and
considering the uncertainties discussed above, we will use the following
value in order to extract $f_K$:
\be \label{eqndelta}
   \frac{\delta \Gamma}{\Gamma_0}(\pi\to\mu\nu_\mu(\gamma))
   = (1.3 \pm 0.2) \%
\ee
Using the experimental input parameters given in the introduction, we
obtain
\be
\label{eqnfk}
   f_K
  = (112.4 \pm 0.9 \pm 0.1) \, \mbox{MeV}
  = (112.4 \pm 0.9) \, \mbox{MeV}
\ee
where the first error is due to $V_{us}$ and the second one to the
radiative correction.

{}From a similar analysis of $\pi_{l2}$ decays, we obtain \cite{Fin94}
\be
\label{eqnfpi}
   f_\pi
  = (92.14 \pm 0.09 \pm 0.09) \, \mbox{MeV}
  = (92.1 \pm 0.1) \, \mbox{MeV}
\ee
(The first error, $\pm 0.09$, is due to $V_{ud}$, and the second one
to the radiative correction.)

This implies
\be
    \frac{f_K}{f_\pi} = 1.22 \pm 0.01
\ee
which agrees with the conventional value \cite{Gas84}.
Note, however, that in our result, the
error is entirely dominated by the uncertainty of $V_{us}$.

\begin{figure}
\caption{Radiative correction to the ratio $R_K $, using
different choices for the hadronic parameters: Standard choice
(I, solid) and variations (II, dashed and III, dotted)}
\label{fig4}
\bildchen{fig4.ps}
{\delta R_K [\%]}
{\mu_{cut}\,[\mbox{GeV}]}
\end{figure}
Let us now come to the prediction for the ratio $R_K$ of the
electronic and muonic decay modes of the kaon. In contrast to $f_K$,
this is a physical observable and therefore free from  ambiguities in
its definition.
In Fig.~\ref{fig4}, we display the
radiative correction $\delta R_K$ of the
ratio using the same standard parameter set (I) and variations (II) and
(III) we used above for $\Gamma(K\to\mu\nu_\mu(\gamma))$. From this, we
obtain the $O(\alpha)$ correction
\be \label{eqnka}
  \delta R_K = - (3.729 \pm 0.023 \pm 0.025) \%
   + O(\alpha^2)
\ee
where the first error ($0.023\%$) gives the matching uncertainty,
estimated by varying $\mu_{cut}$ from $0.75$ up to $3 \, \mbox{GeV}$, and
the second error ($0.025\%$) arises from the uncertainties in the
hadronic parameters.

The error estimate, taking from the dependence on the matching scale
and on the hadronic parameters, actually gives only a lower limit of the
true error, although the very small dependence of the result on the
matching scale supports some confidence in our results.
We will, nevertheless, scrutinize the model dependence further in two
ways:
Firstly we examine, from which scales the contributions to $\delta R_K$
actually come. Secondly we compare the results from our model with
the leading model independent contributions.
\begin{table}
\caption{Contributions from photons with momenta within a given range to
the various radiative corrections}
\label{tabranges}
$$
\begin{array}{|c|rrr|}
\hline \hline
\parbox[c]{5cm}{\str Contribution from photons with $|k_E|$ in the
range [MeV]: \str} &
\multicolumn{3}{c|}{
\parbox[c]{5cm}{{Radiative correction to:}\\
{(all numbers in units of $\%$)}}
}\\
&
    K\to e\nu_e &   K\to\mu\nu_\mu & R_K \\
\hline
0 \dots 125     & -3.926 & -0.515 & 3.411 \\
125 \dots 250   & -0.450 & -0.246 & -0.204\\
250 \dots 500   & -0.201 & -0.115 & -0.086\\
500 \dots 750   & -0.000 &  0.020 & -0.020\\
750 \dots 1000  &  0.047 &  0.055 & -0.008\\
1000 \dots 1500 &  0.121 &  0.126 & -0.005\\
1500 \dots 3000 &  0.322 &  0.318 & 0.004\\
3000 \dots m_Z  &  1.586 &  1.584 & 0.002\\
\hline \hline
\end{array}
$$
\end{table}

Consider Tab.~\ref{tabranges}. We display the contribution to the
radiative corrections from photons with given Euclidean momenta $|k_E|$.
We find that the contributions to the individual decay
rates at large $|k_E^2|$ are quite sizeable. However, the contributions to
the electronic and the muonic mode approach each other for large momenta,
such that the contribution to the correction to the ratio $R_K$ comes
predominantly from very small scales.
Uncertainties from the hadronics in the long distance regime and from
 QCD and wave function corrections in the short distance regime are
large in the intermediate energy range of about $|k_E| = 500 \dots 3000
\,\mbox{MeV}$ only. The total contribution within this range
is given by $0.037 \%$, where we have added the absolute values
in order to take care of cancellations. So by far the largest part of the
radiative correction comes from the region below $500 \, \mbox{MeV}$,
where the model dependence is very small.

\begin{table}
\caption{The different contributions adding up to the total radiative
corrections (the numbers in brackets are obtained assuming exact $SU(3)$
flavour symmetry}
\label{tabeinzeln}
\begin{center}
\begin{tabular}{l c  }
\hline \hline
contribution &
$ \delta R_{\pi}\, [\%]$
\\
\hline \str
\parbox[c]{5cm}{\str (1) effective point meson\\}
& -3.786 \\
\parbox[c]{5cm}{\str (2) vector meson dominance in the point meson
loops \\} &

\begin{tabular}{c} 0.048 \\(0.055)
\end{tabular} \\
\parbox[c]{5cm}{\str (3) hadronic structure dependent loops
proportional to $F_V(0)$\\} & 0.135\\
\parbox[c]{5cm}{\str (4) hadronic structure dependent loops
proportional to $F_A(0)$\\} & -0.134  \\
\parbox[c]{5cm}{\str (5) cutting off the long distance part
at $\mu_{cut} = 1.5 \,\mbox{GeV}$\\}
 & 0.003 \\
\parbox[c]{5cm}{\str (6) short distance corrections\\}
& 0.006 \\
\hline
\parbox[c]{5cm}{\str (7) total\\}
&
\begin{tabular}{c} -3.729 \\ (-3.723) \end{tabular}
\end{tabular} \end{center}
\end{table}
Next we will compare our model with model independent estimates.
In \cite{Ter73} the author calculates the leading logarithmic corrections
to the ratio $R_\pi$, which arise from hadronic structure dependent
effects. He proves that this leading contribution is model independent,
viz.\ independent on the form of the hadronic form factors. The only
assumption needed is that the scale over which the form
factor  vary is given by a large hadronic scale of the order of
$m_\rho$.

The leading correction $\delta R^{(HSD)}$ can be separated
into three contributions
\be
   \delta R^{(HSD)}_K =
   \delta R^{(VMD)}_K + \delta R^{(v)}_K + \delta R^{(a)}_K
\ee
where $R^{(VMD)}_K$ is due to the vector meson dominance of the kaon
electromagnetic form factor, and $R^{(v/a)}_K$ correspond to
virtual corrections proportional to the form factors $F_{V/A}(0)$.
Translating the results to the kaon case, we find
\begin{eqnarray} \label{eqnnum}
   \delta R^{(VMD)}_K & = & \frac{3 \alpha}{\pi}
   \frac{m_\mu^2}{m_\rho^2} \ln \frac{m_\rho^2}{m_\mu^2}
   =  5.2 \cdot 10^{-4}
\nonumber \\
\nonumber \\
   \delta R^{(v)}_K & = & \frac{\alpha}{6 \pi}
   \frac{F_V(0)}{\sqrt{2} m_K f_K}
   \left[ m_K^2 \ln \frac{m_\mu^2}{m_e^2}
   + 4 m_\mu^2 \ln \frac{m_\rho^2}{m_\mu^2}\right]
\nonumber \\
   & = &(14.8 + 1.0) \cdot 10^{-4} =  15.8 \cdot 10^{-4}
\nonumber \\
\nonumber \\
   \delta R^{(a)}_K & = & -\frac{\alpha}{6 \pi}
   \frac{F_V(0)}{\sqrt{2} m_K f_K}
   \left[ m_K^2 \ln \frac{m_\mu^2}{m_e^2}
   + 7 m_\mu^2 \ln \frac{m_\rho^2}{m_\mu^2} \right]
\nonumber \\ \nonumber \\
    & = & -(8.1 + 1.0)\cdot  10^{-4} = - 9.1 \cdot 10^{-4}
\end{eqnarray}
Note that $\delta R_K^{(v)}$ and $\delta R_K^{(a)}$ consist of two
parts.
The first one, being proportional to $m_K^2 \ln (m_\mu^2/m_e^2)$ and
giving rise to a lepton mass singularity, results from contributions to
$C_{lms}$ (see Eqn.~\ref{eqn18}), and the second one, proportional to
$m_\mu^2 \ln (m_\rho^2/m_\mu^2)$, results from the contributions to
$C_2$.

Now let us compare these numbers with the corresponding results
from our model, see Tab.~\ref{tabeinzeln}. In the first row (1) we give
the results obtained with an effective point meson. In the second row
(2) we display the change of the result, when switching on the vector
meson dominance in the meson electromagnetic form factor. In  rows (3)
and (4) we give the contributions from those loop diagrams which
correspond to the SD part in the real radiation. In (2)--(4), we have
extended the loop integration up to $m_Z$, and so in
row (5) we display the change when cutting off the long distance
correction at $\mu_{cut}=1.5\,\mbox{GeV}$, and in row (6) we give
the short distance correction.

Now comparing the model independent numbers $5.2$, $15.8$ and $-9.1$ with
our numbers $4.8$, $13.5$ and $-13.4$ (hadronic structure dependent
corrections in units of $10^{-4}$), we find that the model dependent
contribution in the long distance give rise to an uncertainty
of the order of $\pm 3.9 \cdot 10^{-4}$. This number has been obtained
by adding up quadratically the differences between the model independent
estimates and the results from our model.

Remember that we have included the $SU(3)$ flavour symmetry breaking
in the electromagnetic form factor of the kaon, i.e.\ in row (2) of
Tab.~\ref{tabeinzeln}). But in the vector meson dominance of the
photon coupling in the hadronic structure dependent loops, row (3) and
(4), we used $m_\rho = m_\omega = m_\Phi = m_V = 768\,\mbox{MeV}$.
However, as can be seen from (\ref{eqnnum}), the contribution which depends
on the vector meson mass $m_V$ is very small, $O(1\cdot 10^{-4})$, in both
cases, and the correction is strongly dominated by the modification of
the ratio of lepton mass singularities.
In fact we have checked that the result from our model in rows
(3) and (4) %
changes only by $0.001 \%$ if we use increase $m_V$ up to $1\,
\mbox{GeV}$. So this approximation of flavour symmetry does not induce a
significant uncertainty.

Let us now consider our result for the short distance correction, $0.6
\cdot
10^{-4}$, which includes contributions which depend on
the kaon wave function. However, we find (compare Eqns.~(\ref{eqn23}--%
\ref{eqn24}))
\begin{eqnarray}
   r_K (u=0) & = & 0.6 \cdot 10^{-4} \nonumber \\
   r_K(u=1) & = & 1.0 \cdot 10^{-4}
\end{eqnarray}
So for any choice of the pion wave function, the resulting short distance
correction will be within the range $(0.6 \cdots 1.0 ) \cdot 10^{-4}$.
This should also be compared with
contribution from the leading lepton mass dependent logarithm,
\be
  \frac{2 \alpha}{\pi} \frac{m_\mu^2}{m_\mu^2-\mu_{cut}^2}
  \ln \frac{m_\mu}{\mu_{cut}} = 0.6 \cdot 10^{-4}
\ee
Thus in the short distance part, the model dependent contribution is
within the range
$(0.0 \cdots 0.4)\cdot 10^{-4}$.

And so by adding up all quadratically the  model dependent
corrections, we obtain $ 3.9 \cdot 10^{-4}$, which we consider as an
upper limit for the uncertainty of our result for the $O(\alpha)$
radiative correction.

However, we have to worry about corrections of higher order,
$O(\alpha^n)$. Given the fact that the $O(\alpha)$ correction is
dominated by the contribution from the lepton mass singularity,
$- \frac{3 \alpha}{\pi} \ln \frac{m_\mu}{m_e}$, in \cite{Mar93} the
leading higher order corrections are estimated by summing up all such
logs via the renormalization group,
yielding a correction to $R_K$ of
\be
   1 - \frac{\left(1 - \frac{2 \alpha}{3 \pi} \ln \frac{m_\mu}{m_e}
    \right)^{9/2}}
  {1 - \frac{3 \alpha}{\pi} \ln \frac{m_\mu}{m_e}}
  = 5.5 \cdot 10^{-4}
\ee
And so our final prediction for $\delta R_K$  is
\be
  \delta R_K = -(3.729 \pm 0.039 + 0.055 \pm 0.01) \%
  = (3.78 \pm 0.04) \%
\ee
In the sum, the first number ($-3.729$) is the central value
and the second number ($0.039$)the uncertainty of the $O(\alpha)$ correction.
The third number ($+0.055$) is the leading higher order correction and
$\pm 0.01$ our estimate of the next-to-leading correction.

For the ratio $R_K$ this implies
\be
   R_K = R_K^{(0)} \Big( 1 + \delta R_K \Big)
   = 2.569 \cdot 10^{-5} \times \Big(1 - 0.0378 \pm 0.0004 \Big)
   = (2.472 \pm 0.001) \cdot 10^{-5}
\ee
in good agreement with the particle data group value
$
   R_K = (2.45  \pm 0.11) \cdot 10^{-5}
$ \cite{RPP94}.
It should be noted, however, that this number is based on experiments
published in the the period 1972--1976. Thus there has been no
experimental progress on this number within almost 20 years, a situation
not unusual in kaon physics.
%
\section*{Acknowledgement}
%
The author gratefully acknowledges financial suppurt by the HCM program under
EEC contract number CHRX-CT90026.
%

\end{document}